\documentclass[12pt,letterpaper]{article}%
\usepackage[T2A]{fontenc}
\usepackage{cite}
\usepackage{amsmath}
\usepackage{amsfonts}
\usepackage{amssymb}
\usepackage{graphicx}%
\usepackage{geometry}
\usepackage{fancyhdr, calc, lastpage}       

\newcommand\PACS[1]{\vskip-2.75pc \begin{center}\parbox{.8\textwidth}{\small\bf PACS numbers: \rm #1 \hfill} \end{center}\vskip4pt}%

\begin{document}

\author{A.L. Virovlyansky \\ Institute of Applied Physics RAS, \\
46, Ulyanova St., Nizhny Novgorod, Russia, 603950 \\
 viro@hydro.appl.sci-nnov.ru}
\title{Ray-based analysis of the interference striation pattern in an underwater
acoustic waveguide}
\date{August, 18, 2013}
\maketitle

\begin{abstract}
In some underwater acoustic waveguides with specially selected
sound speed profiles striations or fringes of the interference
pattern are determined by a single parameter $\beta$\ called the
waveguide (or Chuprov) invariant. In the present paper it is shown
that an analytical description of fringes may be possible in a
waveguide with an arbitrary sound speed profile. A simple
analytical expression is obtained for smooth lines formed by local
maxima of the interference pattern. This result is valid at long
enough ranges. It is derived proceeding from a known relation
connecting the differences of ray travel times and the action
variables of ray paths.

\end{abstract}


\section{Formulation of the problem}

Consider a transient wave field excited by a point source in a
range-independent waveguide with the sound speed profile $c(z)$, where $z$ is
the vertical coordinate directed downward. It is assumed that the source
emitting a wideband regular or noise signal $s(t)$ is set at depth $z_{s}$.
Signals arriving at a fixed depth $z=z_{r}$ form function $u(r,t)\,$, where
$r$ is the source-receiver range.

Below we consider two characteristics of the wave field at depth $z_{r}$. One
of them is $\Phi\left(  r,\omega\right)  =\left\vert \tilde{u}(r,\omega
)\right\vert ^{2}$, where $\tilde{u}\left(  r,\omega\right)  =\left(
2\pi\right)  ^{-1}\int d\omega~u(r,t)e^{i\omega t}$. It represents the
interference pattern in the range-frequency plane $\left(  r,\omega\right)  $.
Local maxima and minima of $\Phi\left(  r,\omega\right)  $ form smoth curves
which we call the interference lines and denote $\omega\left(  r\right)  $.
The interference pattern consists of fringes (or striations) localized in the
vicinities of the interference lines
\cite{Chuprov82,JKPS2011,Harrison2011,Dspain99,Schmidt2010,Krolik2008,Rouseff2012}%
.

Another characteristic of the wave field is represented by the autocorrelation
function
\begin{equation}
K\left(  r,\tau\right)  =\int dt~u(r,t+\tau)u^{\ast}\left(  r,t\right)
=2\pi\int d\omega~\Phi\left(  r,\omega\right)  e^{-i\omega\tau}.\label{K-Phi}%
\end{equation}
It determines the interference pattern in the range--time delay plane $\left(
r,\tau\right)  $. In this plane also there are fringes localized in the
vicinities of the interference lines $\tau\left(  r\right)  $.

In some waveguides with specially selected profiles $c(z)$ -- we will call
them the Chuprov waveguides -- the interference lines are determined by simple
equations%
\begin{equation}
\frac{r}{\tau}\frac{d\tau}{dr}=-\beta,\;\frac{r}{\omega}\frac{d\omega}%
{dr}=\beta,\label{Chuprov-eq}%
\end{equation}
where $\beta$ is the so-called waveguide (or Chuprov) invariant
\cite{Chuprov82,JKPS2011,Harrison2011}. In a Chuprov waveguide $\beta$ is the
same constant for all the interference lines. Then Eqs. (\ref{Chuprov-eq}) are
readily solved to yield
\begin{equation}
\tau(r)=~Cr^{-\beta},\omega(r)=C_{1}r^{\beta},\label{IL}%
\end{equation}
where $C$ and $C_{1}$ are some constants. Our objective in the present work is
to derive an analytical expression for the interference line in a waveguide
with an arbitrary sound speed profile.

\section{Equation for the interference line}

For solving this problem we will use the ray representation of the wave field
\cite{JKPS2011,BL2003}. In the geometrical optics approximation the signal at
observation point $\left(  r,z_{r}\right)  $ is presented in the form%
\begin{equation}
u(r,t)=\sum_{n}u_{n}\left(  r,t-t_{n}\right)  ,\label{u-un}%
\end{equation}
where $t_{n}$ is the travel time of the $n$-th eigenray, $u_{n}\left(
t\right)  $ is the sound pulse coming through this eigenray. The eigenrays are
numbered in such a way that the same subscript $n$ at different ranges $r $ is
associated with eigenrays with the same identifier $\left(  N,\mu,\nu\right)
$, where $N$ is the number of ray lower turning points (number of cycles),
$\mu=\pm1$ and $\nu=\pm1$ determine the signs of the ray grazing angles at the
end points \cite{Harrison2011}.

According to Eq. (\ref{u-un}), the autocorrelation function of the received
sound signal is%
\begin{equation}
K(r,\tau)=\sum_{n,m}q_{nm}(r,\tau-\tau_{nm}),\label{K}%
\end{equation}
where $q_{nm}\left(  r,\tau\right)  =\int dt~u_{n}(r,t+\tau)u_{m}^{\ast
}\left(  r,t\right)  $,~$\tau_{nm}=t_{n}-t_{m}$. At a fixed range $r$, each
term $q_{nm}\left(  r,\tau\right)  $ represents a peak with maximum at
$\tau=0$ and width $O\left(  1/\Delta f\right)  $, where $\Delta f$ is the
bandwidth of the emitted signal $s(t)$.

It follows from Eqs. (\ref{K-Phi}) and (\ref{K}) that%
\begin{equation}
\Phi\left(  r,\omega\right)  =\sum_{n,m}\tilde{q}_{nm}\left(  r,\omega\right)
e^{i\omega\tau_{nm}(r)},\;\label{Phi-q}%
\end{equation}
where $\tilde{q}_{nm}\left(  r,\omega\right)  =\left(  2\pi\right)  ^{-2}\int
d\tau~q\left(  \tau\right)  e^{i\omega\tau_{nm}(r)}$. According to Eqs.
(\ref{K}) and (\ref{Phi-q}), the interference patterns in both planes $\left(
r,\tau\right)  $ and $\left(  r,\omega\right)  $ to a significant extent are
determined by functions $\tau_{nm}(r)$.

Let us divide all the eigenrays arriving at the observation point
$\left( r,z_{r}\right)  $ in groups of fours. Each group includes
eigenrays whose idenifiers $\left(  N,\mu,\nu\right)  $ have the
same value of $N$ and different pairs $(\mu,\nu)$
\cite{Harrison2011}. If the source and receiver are located at
relatively small depths -- this case will be considered in the
rest of this paper -- the travel times of eigenrays belonging to
the same group of four are close. The spread of these travel times
is small compared to the difference between travel times of
eigenrays from different groups. In Eq. (\ref{K}) each pair of
groups of four is presented by 16 terms which may strongly
overlap. Their superposition form a fringe in the plane $\left(
r,\tau\right)  $ located in the vicinity of the interference line
$\tau\left( r\right)  =\tau_{nm}\left(  r\right)  $, where
$\tau_{nm}$ is the difference between travel times of two
eigenrays taken from two different groups. The value of
$\tau_{nm}$ weakly depends on the choice of particular eigenrays
taken from groups forming the fringe.

An approximate analytical expression for the interference line
$\tau(r)$ can be derived proceeding from the known relation
connecting the difference in ray travel times and the action
variables of the ray paths
\cite{Harrison2011,V85,V95,MW83,Vbook2010}. Assume that the sound
speed profile $c\left(  z\right)  $ has a single minimum at depth
$z=z_{a}$. Then the action variable of a ray path intersecting the
horizon $z_{a}$ at a grazing angle $\chi$ is \cite{Vbook2010}
\begin{equation}
I=\frac{1}{\pi}\int_{z_{\min}}^{z_{\max}}dz~\sqrt{c_{0}^{2}/c^{2}\left(
z\right)  -\cos^{2}\chi},\label{I}%
\end{equation}
where $c_{0}=c(z_{a})$, $z_{\min}$ and $z_{\max}$ are the upper and lower ray
turning depths, respectively. The cycle length of the ray path,%
\begin{equation}
D=2\cos\chi\left.  \int_{z_{\min}}^{z_{\max}}dz\right/  ~\sqrt{c_{0}^{2}%
/c^{2}\left(  z\right)  -\cos^{2}\chi},\label{D}%
\end{equation}
can be considered as a function of the action $I$. We denote this function
$D(I)$.

Take two eigerays connecting the source and receiver located at the same
depth. We assume that the eigenrays have exactly $N+\Delta N$ and $N$ cycles
of oscillations. The travel times of these eigenrays denote $t_{N+\Delta N}$
and $t_{N}$, respectively. Similarly, their action variables denote
$I_{N+\Delta N}$ and $I_{N}$. The latter are determined by the relations%
\begin{equation}
D(I_{N+\Delta N})=r/(N+\Delta N),\;D(I_{N})=r/N.\label{DN}%
\end{equation}

On condition that%
\begin{equation}
N\gg\Delta N,\label{N-large}%
\end{equation}
$I_{N+\Delta N}$ and $I_{N}$ are close and the difference between the eigenray
travel times is given by the approximate relation \cite{Vbook2010,V2007a}
\begin{equation}
\tau=t_{N+\Delta N}-t_{N}=2\pi I\Delta N/c_{0},\label{main}%
\end{equation}
where $I=\left(  I_{N+\Delta N}+I_{N}\right)  /2$. This relation was derived
by different authors \cite{Harrison2011,V85,V95,MW83}. The relationship
between the ray travel times and action variables is studied in detail in
monograph \cite{Vbook2010}. The monograph provides a detailed dervation of Eq.
(\ref{main}) and its generalizations.

If the distance between the source and receiver changes by a small amount
$\delta r$, then the actions $I_{N+\Delta N}$ and $I_{N}$ change by $\delta
I_{N+\Delta N}$ è $\delta I_{N}$, respectively. It follows from Eqs.
(\ref{DN}) and (\ref{N-large}) that the mean value of action variables $I$
changes by%
\begin{equation}
\delta I=\frac{1}{2}\left(  \delta I_{N+\Delta N}+\delta I_{N}\right)
\simeq\frac{D\left(  I\right)  }{D^{\prime}\left(  I\right)  }\frac{\delta
r}{r},\label{del_I}%
\end{equation}
where $D^{\prime}(I)=dD(I)/dI$. The corresponding change in the difference of
travel times $\tau$ denote $\delta\tau$. According to Eq. (\ref{main}),
$\delta\tau/\tau=\delta I/I$. Substiting Eq. (\ref{del_I}) in this relation
yields the desired equation for the interference line%
\begin{equation}
\frac{r}{\tau}\frac{d\tau}{dr}=-\beta\left(  I\right)  ,\label{tau-eq}%
\end{equation}
where%
\begin{equation}
\beta\left(  I\right)  =-\frac{D\left(  I\right)  }{ID^{\prime}\left(
I\right)  }.\label{b-I}%
\end{equation}
Equation (\ref{b-I}) relating the Chuprov parameter $\beta$ with the action
variable of the ray path was obtained earlier in Ref. \cite{BB2009}.

In the Chuprov waveguide $D(I)$ is a power function. Then $\beta$
does not depend on $I$ and has the same value for all the
interference lines. An
example is given by a waveguide with the sound speed profile%
\begin{equation}
c(z)=c_{0}\sqrt{1+\left\vert \frac{z-z_{a}}{a}\right\vert ^{g}},\label{c-g}%
\end{equation}
where $c_{0}$, $a$ and $g$ are constants. For relatively flat rays
in such a waveguide $D(I)\sim I^{(2-g)/(2+g)}$ and we find
$\beta=(g+2)/(g-2)$ \cite{Harrison2011}. If $g=1$ formula
(\ref{c-g}) determines a waveguide with the squared index of
refraction represented by a linear function of $z$. Then
$\beta=-3$ \cite{Chuprov82,JKPS2011,Harrison2011}. Another
well-known result is obtained for $g=\infty$. In this case
$c(z)=c_{0}$ and we arrive at the Pekeris waveguide where
$\beta=1$ \cite{Chuprov82,JKPS2011,Harrison2011}.

Consider a fringe in a waveguide with an arbitrary sound speed profile formed
by a pair of groups of four with $N+\Delta N$ and $N$ cycles of oscillation.
Using Eq. (\ref{main}), replace $I$ on the right-hand side of Eq. (\ref{b-I})
by $\tau c_{0}/(2\pi\Delta N)$. This yields an equation for $\tau$ which is
readily solved. It turns out that the interference line $\tau(r)$ which takes
value $\tau_{0}$ at range $r_{0}$ is determined by the relation
\begin{equation}
\frac{D\left(  \frac{c_{0}\tau}{2\pi\Delta N}\right)  }{D\left(  \frac
{c_{0}\tau_{0}}{2\pi\Delta N}\right)  }=\frac{r}{r_{0}}.\label{main-eq}%
\end{equation}
According to this formula, the shape of the interference line, generally,
depends on $\Delta N$. Our assumption that the source and receiver have the
same depths is made only to simplify the derivation of (\ref{main-eq}). This
result remain valid for $z_{r}\neq z_{s}$. It should be emphasized that Eqs.
(\ref{main}) and (\ref{main-eq}) are valid not only for eigenrays with turning
points within the water bulk, but for eigenrays reflected from the surface
or/and bottom, as well.

In a Chuprov waveguide Eq. (\ref{main-eq}) translates to%
\begin{equation}
\frac{\tau}{\tau_{0}}=\left(  \frac{r}{r_{0}}\right)  ^{-\beta}%
.\label{Chuprov-line}%
\end{equation}
As should be, Eq. (\ref{Chuprov-line}) coincides with first of Eqs. (\ref{IL}).

Generally, a fringe in plane $\left(  r,\tau\right)  $ located in the vicinity
of an interference line described by Eq. (\ref{main-eq}) can be observed only
if terms $q_{nm}$ forming the fringe do not overlap with terms forming other
fringes. But in the Chuprov waveguides this requirement is \textbf{not
necessary}. The point is that the function $\tau_{nm}(r)$ in the Chuprov
waveguide has the following property. If at range $r_{0}$ function $\tau
_{nm}\left(  r\right)  $ takes value $\tau_{0}$, then at other ranges $r$
(both at $r>r_{0}$ and $r<r_{0}$) this function, according to Eqs. (\ref{IL})
and (\ref{Chuprov-line}), is complitely determined by $\tau_{0}$ and does not
depend on other eigenray parameters. Therefore, any local extremum of function
$K(r,\tau)$ varies with $r$ according to Eq. (\ref{Chuprov-line}) even if it
is formed by overlapping terms of sum (\ref{K}). For the same reason, local
extrema of function $\Phi(r,\omega)$ in the Chuprov waveguides form smooth
lines%
\begin{equation}
\omega(r)=\pi M/\tau(r),\label{om-tau}%
\end{equation}
where $M$ is an integer, even if terms in sum (\ref{Phi-q}) overlap
\cite{Harrison2011}.

Formula (\ref{main-eq}) is the main result of this work. It
generalize the Chuprov equations (\ref{IL}) to a waveguide with an
arbitrary sound speed profile.

\section{Numerical example}

To illustrate the applicability of Eq. (\ref{main-eq}) in a non-Chuprov deep
water waveguide consider the sound speed profile shown in the left panel of
Fig. 1. The right panel of Fig. 1 presents the dependence of parameter $\beta$
determined by Eq. (\ref{b-I}) on the ray grazing angle at the sound channel
axis $z_{a}=0.7$ km. It is seen that the values of $\beta$ are significantly
different for different rays. Since at large depths the sound speed is a
linear function of $z$, it is not surprising that the values of $\beta$ for
steep rays are close to $-3$.

\begin{figure}[ptb]
\centering
\includegraphics[
height=6.4cm, width=8.9cm ]{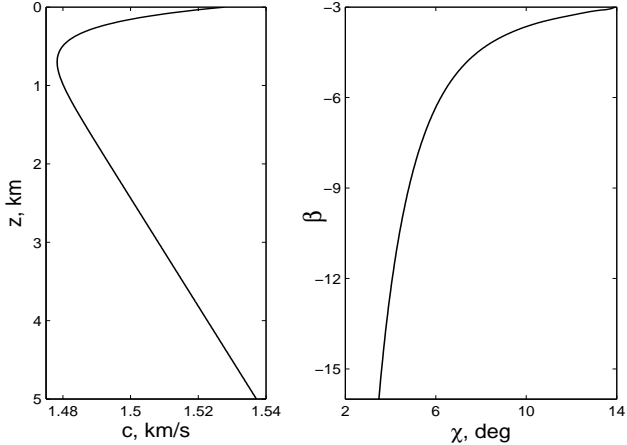}\caption{Left panel: sound
speed profile used in numerical simulation. Right panel: parameter
$\beta$ determined by Eq. (\ref{b-I}) vs. ray grazing angle $\chi$
at the sound channel axis $z_{a}=0.7$
km.}%
\end{figure}

Using a standard mode code we evaluated a sound field emitted by a point
source set at range $r=0$ and depth $z_{s}=0.3$ km and radiating a sound pulse
$s(t)=\exp\left(  -\pi t^{2}/T^{2}-2\pi if_{0}t\right)  $, where the pulse
length $T=0.002$ s and the central frequency $f_{0}=250$ Hz. Only modes with
turning points within the water bulk were taken into account.

We analyse sound pulses at points located at depth $z_{r}=z_{s}$ within the
range interval from $r=212$ km to $r=312$ km. Signals $u(r,t)$ arriving at
these points without reflections from the waveguide boundaries are formed by
groups of four eigenrays with $N=4,5,\ldots,8$. In the plot of function
$\left\vert u(r,t)\right\vert $ shown in Fig. 2 each group of four manifests
itself as a fringe. A number next to each fringe indicates the parameter $N$
of eigenrays contributing to the fringe. Since the source and receivers in our
example are set at the same depth, in each group of four there are two
eigenrays with equal travel times. Therefore fringes corresponding to some
groups (with $N=$ 4, 5, and 6) are split into three (not four) more narrow fringes.

\begin{figure}[ptb]
\centering
\includegraphics[
height=8.4cm, width=11.9cm ]{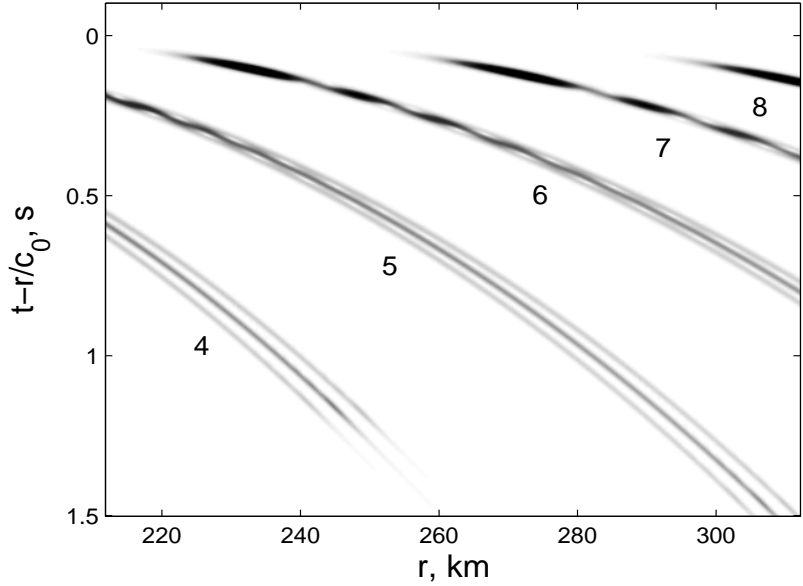}\caption{Wave field amplitude
$\left\vert u(r,t)\right\vert $ at depth $z_{r}=0.3$ km. The wave field is
excited by a point source set at depth $z_{s}=z_{r}$ and emitting the short
sound pulse $s\left(  t\right)  $ described in the text. The time is reckoned
from $r/c_{0}$, which is the arrival time of an axial ray. Five fringes formed
by groups of four eigenrays with $N=4,5,...,8$ are clearly seen.}%
\end{figure}

The interference pattern presented by the autocorrelation function $K(r,\tau)
$ is shown in Fig. 3. The fringes formed by pairs of interfering groups of
four eigenrays are clearly seen in this plot. A pair of numbers next to each
fringe indicates parametes $N$ corresponding to the groups of four forming the
fringe. White dashed curves graph interference lines predicted by Eq. (\ref{main-eq}%
). Parameters $r_{0}$ and $\tau_{0}$ used in the evaluation of a white curve
are the coordinates of a point selected somewhere near the center of the
corresponding fringe.

\begin{figure}[ptb]
\centering
\includegraphics[
height=8.4cm, width=11.9cm ]{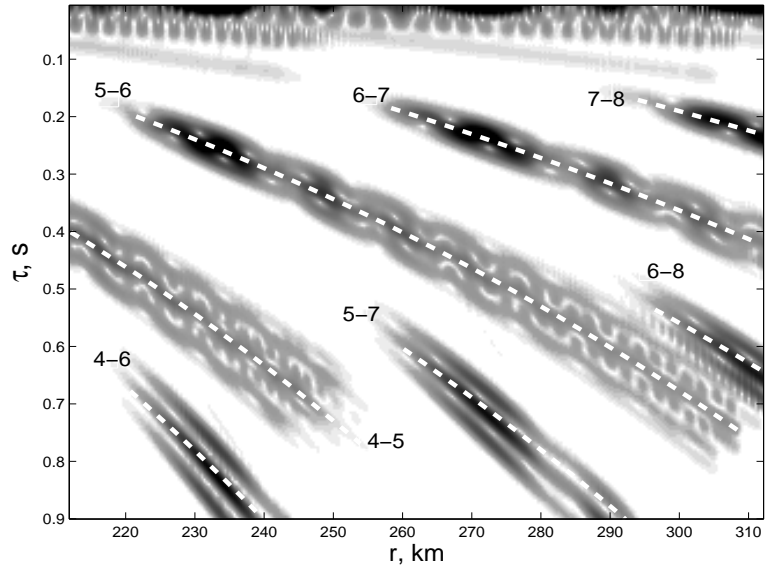}\caption{Interference pattern
represented by the autocorrelation functions of signals shown in Fig. 2. Seven
fringes formed by pairs of groups of four are resolved. Two numbers next to
each fringe indicate the numbers of cycles $N$ for a corresponding pair of
groups of four. White dashed curves represent interference lines predicted by
Eq. (\ref{main-eq}).}%
\end{figure}

The applicability of Eq. (\ref{main-eq}) requires the closeness of the action
variables of rays forming the interference line. This requirement is satisfied
only at long enough ranges where condition (\ref{DN}) is met. Formula
(\ref{main}) is the most accurate for $\Delta N=1$. Figure 3 presents fringes
corresponding to $\Delta N=1$ and $\Delta N=2$. Consistent with our
expectation, the interference lines predicted by Eq. (\ref{main-eq}) for
$\Delta N=1$ better describe the behavior of fringes than the lines
corresponding to $\Delta N=2$.

Observation of fringes associated with the interference lines predicted by Eq.
(\ref{main-eq}) in a non-Chuprov waveguide can be a difficult task. This task
is especially complicated in the $\left(  r,\omega\right)  $ plane, where,
according to Eq. (\ref{om-tau}), each group of four produces a whole set of
fringes. The use of a vertical antenna may simplify the resolution of
individual fringes. It allows one to diminish the number of terms in sums
(\ref{K}) è (\ref{Phi-q}) by selecting waves propagating at grazing angles
within a narrow interval. However, the discussion of this issue is beyond the
scope of the present paper.

\section*{Acknowledgments}

The work was supported by the Program "Fundamentals of acoustic diagnostics of
artificial and natural media" of Physical Sciences Division of Russian Academy
of Sciences, the Grants Nos. 13-02-00932, 13-02-97082 and 13-05-90307 from the
Russian Foundation for Basic Research, the Leading Scientific Schools grant
No. 333.2012.2.

\end{document}